\documentclass[fleqn,10pt]{wlscirep}
\usepackage[utf8]{inputenc}
\usepackage[T1]{fontenc}

\usepackage{graphicx}
\usepackage{dcolumn}
\usepackage{bm}
\usepackage{acronym}
\newacro{QML}{Quantum Machine Learning}
\newacro{LHC}{Large Hadron Collider}
\newacro{ML}{Machine Learning}
\newacro{HEP}{High Energy Physics}
\newacro{QSVM}{Quantum Support Vector Machines}
\newacro{SVM}{Support Vector Machine}
\newacro{QNN}{Quantum Neural Networks}
\newacro{MC}{Monte Carlo}
\newacro{SM}{Standard Model}
\newacro{BSM}{Beyond Standard Model}
\usepackage{hyperref}
\usepackage{xcolor}
\hypersetup{
    colorlinks=true,
    linkcolor=black,
    filecolor=black,      
    urlcolor=black,
    citecolor=black,
}
\usepackage[style=ieee, citestyle=numeric-comp, sorting=none, backend=biber]{biblatex}
\usepackage{csquotes}

\let\citen=\cite
\let\cite=\supercite

\title{Machine Learning for Anomaly Detection in Particle Physics}

\author[1,*]{Vasilis Belis}
\author[1,*]{Patrick Odagiu}
\author[1,*]{Thea Kl\ae boe Aarrestad}

\affil[1]{Institute for Particle Physics and Astrophysics\\
ETH Z\"urich, 8093 Z\"{u}rich, Switzerland
}

\affil[*]{e-mail: vbelis@ethz.ch, podagiu@ethz.ch, thea.aarrestad@cern.ch}

\begin{abstract}
The detection of out-of-distribution data points is a common task in particle physics. It is used for monitoring complex particle detectors or for identifying rare and unexpected events that may be indicative of new phenomena or physics beyond the Standard Model. Recent advances in Machine Learning for anomaly detection have encouraged the utilization of such techniques on particle physics problems. This review article provides an overview of the state-of-the-art techniques for anomaly detection in particle physics using machine learning. We discuss the challenges associated with anomaly detection in large and complex data sets, such as those produced by high-energy particle colliders, and highlight some of the successful applications of anomaly detection in particle physics experiments.
\end{abstract}

\begin{document}

\flushbottom
\maketitle
\tableofcontents
\thispagestyle{empty}

\section{Introduction}

Anomaly detection plays an important role in various scientific disciplines, aiding in the discovery of rare and unusual events that deviate significantly from the norm. I the context of high energy physics (HEP), two primary types of anomaly detection are used: outlier detection and finding over-densities. Outlier detection focuses on the identification of unusual or unexpected events that stand out from the norm. These outliers are typically found in the tails of distributions, representing rare occurrences. In HEP, the search for outliers becomes particularly crucial as it unveils exceptional phenomena or anomalies that hold valuable insights into the fundamental nature of particle interactions. Detecting over-densities in data involves a slightly different approach to anomaly detection. This type of anomaly detection can be seen as analogous to the traditional bump hunt, where one looks for a localized excesses in data points compared to the expected distribution. These over-densities, or resonances, may indicate the presence of new particles or unexpected physical processes. Traditional rule-based approaches for anomaly detection can be limited by the complexity and variability of the data, making it challenging to define rules that cover all possible scenarios. As a result, machine learning (ML) techniques have gained popularity as a more flexible and powerful approach for performing anomaly detection~\cite{noveltydetection}.

ML-based anomaly detection methods have especially been gaining popularity in particle physics as a way of extracting potential new physics signals in a model-agnostic way, by rephrasing the problem as an out-of-distribution detecting task~\cite{Kasieczka_2021}. In this context, model-agnostic refers to assuming no, or at least minimal, prior information regarding the physical model describing the new-physics phenomena. A typical search for new physics signatures involves looking for a specific signal and maximizing the analysis sensitivity for that single model. This analysis is not useful to investigate other new physics models. In an anomaly detection driven search, however, the aim is to be model-agnostic and only look for deviations from the background. This is less sensitive to any model that is biased to a specific signal, yet, it enables the simultaneous search of multiple new physics scenarios. An additional advantage of anomaly detection is that it allows algorithms to undergo direct training on unlabeled data. This has generated substantial interest in the physics community, resulting in several community challenges on anomaly detection driven searches for new physics~\cite{Kasieczka:2021tew,Kasieczka_2021,ADC,darkmachines}.

This review paper aims to provide an overview of the various machine learning techniques used for anomaly detection in particle physics, including their strengths and limitations. We will focus on its usage in high energy particle physics, but several of the techniques discussed generalize to neutrino physics, astro-particle physics, and gravitational wave detection. Our main focus will be on the usage of anomaly detection as a means of discovering new physics, but we will also discuss its usage in system monitoring. A more exhaustive summary of outlier detection can be found in Refs.~\citen{adsurvey,10.5555/3086742}. This review is organized as follows. First, we will give a brief introduction to training methodologies in Section~\ref{sec:paradigms}. This is followed by an overview over how anomaly detection is used for model-independent searches in high energy physics experiments in Section~\ref{sec:lhc_searches}, including recent results. In Section~\ref{sec:realtime} we discuss anomaly detection for triggering and in Section~\ref{sec:dqm} we briefly discuss usecases in detector monitoring. Finally, in Section~\ref{sec:qml}, we discuss anomaly detection in the emerging field of quantum machine learning.

\section{Training paradigms in particle physics: Degree of supervision}
\label{sec:paradigms}

Particle physics data is unique in that it inherently cannot be labeled in the same way that, for instance, images can. Fundamentally, every data sample can be a signal process, a background process or a quantum mechanical superposition of the two. Consequently, typical deep learning setups where a loss function defined in terms of predicted versus true labels, and which is minimized over some dataset containing such true labels, is impossible. In order to account for this, highly accurate simulation of physical processes exist. This simulated data acts as a labeled surrogate of the real processes measured in particle physics experiments.

For the training of deep neural networks in a \textit{supervised manner} in particle physics, simulated labeled data must be used. To utilize the vast amount of unlabeled data, and overcome difficulties related to differences between simulation and data, other strategies such as \textit{weakly- or semi-supervised} training paradigms have been developed. In semi-supervised learning, a small amount of labeled data is combined with a large amount of unlabeled data at training time. In weakly-supervised learning, noisy or imprecise sources are used to label the training data. One can also take advantage of \textit{self-supervised} methods. Here, one takes advantage of underlying structure in the data to obtain supervisory signals from the data itself. This is for instance the case for an autoencoder, a model that is trained to compress and decompress the input, and uses the difference between the original and the reconstructed input to compute a reconstruction error. The original input then serves as the label that the reconstructed input is trying to target. Self-supervised and \textit{un-supervised} methods have been gaining significant popularity in particle physics, not only due to the large amount of unlabeled data available, but also as a way of reducing the high degree of model-dependence introduced by using simulated data do define search strategies. These training paradigms are increasingly applied to natural language processing and computer vision, aimed at harnessing the growing reservoir of unlabeled data accessible on the internet. This surge has resulted in the emergence of innovative and effective approaches, which can be adapted and employed within the physical sciences. In the following, we will mainly focus on self-, semi- and weakly- supervised methods although supervised approaches to anomaly detection in particle physics also exist~\cite{antiqcd}.

\section{Anomaly detection for model-agnostic new physics searches}
\label{sec:lhc_searches}
Searching for physics beyond the Standard Model is one of the most important aspects of the physics program at the Large Hadron Collider (LHC). Since the start of proton-proton collisions at the LHC in 2011, the ATLAS~\cite{ATLAS} and CMS~\cite{CMS} Collaborations have derived stringent bounds on a range of new physics signatures, pushing the allowed mass range for many postulated new particles far into the TeV scale. While it is possible that these particles have yet to be observed because they are too heavy to be produced at the LHC, or have to small cross section to be detected with the current data size, it could also be that new particles are kinematically accessible and produced at observable rates, but our current methods of detection prevent their discovery.

Searches for new physics processes at particle colliders are usually performed as \textit{blind searches}. Such searches proceed by defining a region of interest in the parameter space, using simulated data of the signal and the Standard Model background processes in order to enhance the data purity. The data is only looked at in the very end where it is tested for the presence of signal through a simultaneous fit of the signal and background probability distributions, hoping to extract a non-zero signal component.

Hundreds of such searches have been performed for hundreds of different potential new particles, but thus far none have been discovered. Despite this, there are still regions of the data that have not yet been probed for the presence of a signal. This has led to an increased interest in more \textit{model-agnostic} search strategies. Model-independent searches is nothing new in high energy particle physics, and strategies relying less on a signal hypothesis have been devised and utilized~\cite{D0:2000vuh,H1:2008aak,H1:2004rlm,Cranmer:823591,CDF:2007iou,CDF:2007ykt,CDF:2008voc,CMS-PAS-EXO-14-016,CMS-PAS-EXO-10-021,CMS-PAS-EXO-19-008,CMS:2020zjg,ATLAS:2018zdn,ATLAS-CONF-2014-006,ATLAS-CONF-2012-107,ATLAS:2020iwa}.
These mainly take advantage of Monte Carlo simulation, and use this to compare distributions in the observed data to simulation across several observables and many histogram bins. The drawback of this methodology is that one needs to rely on accurate simulation, and also that, due to the vast size of the parameter space being searched, an observation that appears statistically significant could potentially be the result of a statistical fluctuation.

In the following, we discuss machine learning techniques which mitigate some of these challenges and have the potential to improve and extend model-independent searches.

\subsection{Overdensity estimation}

In order to train the most powerful ML-based classifier to discriminate signal from background, one would ideally train a network in a supervised manner with labeled data. This relies on a signal hypothesis that is chosen a-priori. An early attempt at discriminating background from "everything else" in order to obtain some degree of model-independence, was demonstrated in Ref.~\citen{antiqcd}. Targeting searches for new physics in hadronic final states, a classifier was trained to discriminate QCD jets from various potential signal jets using Monte Carlo simulation. The disadvantage of such an approach is the dependence on signal simulation and which signals are to be included in the training.
Although simulated particle physics data is highly accurate over several orders of magnitude in length scale, simulation is known to not fully accurately reproduce collider data and this disagreement affects the tagging performance. Using weakly- or self-supervised (see Section~\ref{sec:selfsupervised}) methods, algorithms can be trained directly on the data itself which has the added benefit of not having to derive transfer factors when training on synthetic data and testing on real data.

\subsubsection{Weakly supervised methods}
\begin{figure}
  \centering
  \includegraphics[width=0.50\textwidth]{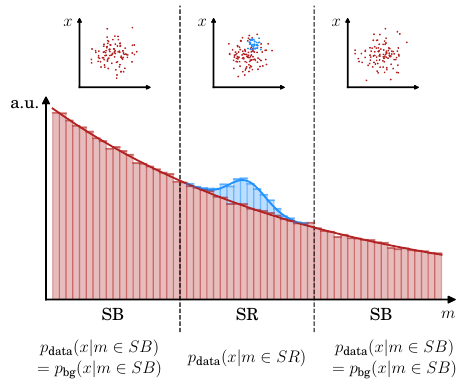}
  \caption{Weakly supervised density estimation techniques like CWola, CATHODE and Tag'N'Train take advantage of the fact that the signal (blue) can be localized in a specific region of phase space (SR). One can then use data sidebands (SB) to either estimate the background distribution (red) in the signal region, or to train a weakly supervised classifier (CWoLa) between SR and SB (figure from Ref.~\citen{Hallin:2021wme}).}
  \label{fig:fig1}
\end{figure}
In weakly supervised learning, impure or noisy data sources can be used to label signal and background data in such a way that models can be trained in a supervised manner. Such methods can be utilized for anomaly detection when the signal is unknown, but there exist datasets where both signal and background are expected to be present in some relative fraction. This can be achieved by placing weak assumptions on the signal and background processes using domain knowledge.

The goal of the weakly supervised methods we will discuss here, is to learn an approximation of the likelihood ratio $R(x)$ between the underlying probability densities of background $p_\text{bg}(x)$ and data (possibly including signal) $p_\text{data}(x)$, as a function of some input variables $x$:
\begin{equation}\label{eq:weak-supervision-likelihood-ratio}
R(x)=\frac{p_\text{data}(x)}{p_\text{bg}(x)}.
\end{equation}
This likelihood ratio, if it could be learned exactly, would be the most powerful model-agnostic anomaly detector, as given 
\begin{equation}
p_\text{data}(x)=(1-\epsilon)p_\text{bg}(x)+\epsilon p_\text{sig}(x),
\end{equation}
where $p_\text{sig}(x)$ is the probability density of signal, it would be monotonically related to the signal-to-background LR for any signal present in the data. A strategy for learning a good approximation of the likelihood ratio $R(x)$, is to train a classifier between data from a signal enriched region and samples drawn from a (fully data-driven) background model. If the background model is accurate and the classifier is well-trained, this approaches the likelihood ratio $R(x)$ by the Neyman-Pearson Lemma~\cite{neyman1933ix}.

Hence, the aim is to test whether the signal region data contains a combination of signal and background data. In the event that there are signal events present in the signal region, the classifier can differentiate between the signal region data and the background template. The true signal events are expected to have higher classification scores than the true background data. A cut on this classifier score can then be used to enhance the significance of signal events, making it a useful anomaly detection metric.

 In Ref.~\citen{Dery2017WeaklySC} a method referred to as Learning from Label Proportions~\cite{LLP} was utilized to discriminate between quarks and gluons using impure data samples. Despite not having access to the per-instance labels, the class proportions could be derived using domain knowledge. A supervised task was then defined using the class proportions themselves as the target, although operating the algorithm at a per-instance level. This concept has been extended in in the Classification WithOut Labels (CWola)~\cite{cwola} framework. In this setup, the class proportions themselves do not need to be known, and it is enough to have two datasets at hand with an unequal fraction of signal instances in each set. A standard classifier can then be trained to discriminate between the two mixed datasets, and this can be shown to be the optimal classifier to discriminate between signal and background instances. The larger the difference in signal fraction between is dataset, the better the classifier becomes. The challenge is being able to design such mixed datasets, especially for a model-independent setup.

\begin{figure}
  \centering
  \includegraphics[width=0.50\textwidth]{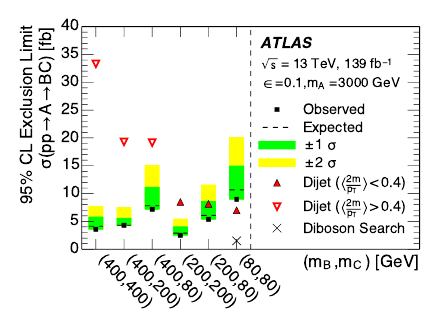}
  \caption{95\% confidence level upper limits on the crossection for a wide range of different signal models using the CWola bumphunt method~\cite{ATLAS:2020iwa}.}
  \label{fig:atlascwola}
\end{figure}

The CWola strategy has been demonstrated and deployed for various model-independent search setups. In Ref.~\citen{cwolabumphunt}, the authors introduce the \textit{CWola bumphunt}. In this setup, one attempts to look for new, heavy generic particles that resonate around the particle mass in the dijet invariant mass spectrum. Starting from the weak assumption that this is a localized, narrow resonance, two mixed samples are created in the following way: The region in the dijet invariant mass close to the particle mass is defined as the signal-enriched mixed sample, and the regions next to it are defined as background-enriched regions. This is illustrated in Fig.~\ref{fig:fig1}. In this way, the dijet invariant mass sideband regions serve as the background samples; these can serve as a good model for the background if the input features are statistically independent from the dijet invariant mass. If there is a signal present in the signal-like region around the particle mass, the classifier learns to identify it, while in the absence of a signal the classifier will likely learn random noise as there would be no difference between the two groups of events. It is crucial that the features being used for classification are not correlated with the dijet invariant mass. Otherwise, the classifier will be able to differentiate background events in the signal region from those in the adjacent dijet invariant mass regions used as the background-enriched mixed sample. Background events within the signal region will then be classified as signal-like, which can introduce artificial sculpting of the dijet invariant mass distribution. Note that the above strategy only works for narrow resonances, if there is a significant amount of signal in both datasets, as would be the case for a broad resonance, the classification performance is reduced. This method was used to analyze data collected by the ATLAS experiment in the search for generic new heavy resonances decaying into jets in Ref.~\citen{ATLAS:2020iwa}, a first of its kind using weak supervision for model-agnostic searches. The power of this analysis can be seen in Figure~\ref{fig:atlascwola}. This plot shows 95\% confidence level upper limits on the crossection for a wide range of different signal models. The results are compared to those of a generic dijet search and to dedicated searches, when these exist. This demonstrates that utilizing powerful anomaly detection techniques like CWola, one can very efficiently search for a wide range of potential New Physics scenarios using a single method.

This methodology can also be applied in other setups than for a dijet bumphunt. In Ref.~\citen{cwola_monojet}, model-agnostic learning using the CWola method is harnessed in order to improve the sensitivity of searches for new physics models with anomalous jet dynamics and a mono-jet signature. Focusing on cases where a heavy new particle decays into two jets which hadronize partially in the dark sector (making them \textit{semi-visible jets}), and where one of the jets become completely invisible and the other partially visible, anomaly detection is utilized to detect the semi-visible anomalous jet. The degree of visibility of this jet can vary, making it difficult to train a supervised algorithm for each visibility fraction. Rather, CWola is deployed to train a generic anomalous jet identification classifier. The dominant background for a mono jet search is the electroweak production of vector bosons and jets, where the vector boson further decays to neutrinos, $Z(\nu\nu)$+jets. The experimental signature is missing transverse energy and a jet, mimicking the signal signature. Taking advantage of the fact that the vector boson also can decay visibly into two leptons and in these cases the jet remains the same, a background enriched control CWola sample can be defined using $Z(\ell\ell)$+jets events. None of the signal should be present in events with a di-lepton and jet signature. A model-independent anomalous jet tagger is then trained supervised to discriminate between jets coming from a $(\ell\ell)+jet$ and a $(\nu\nu)$+jet sample. If the monojet signature is present, CWola guarantees that the best algorithm trained to distinguish between these two regions, is also the best algorithm to discriminate between a normal SM jet from the V+jet background, and a semi-visible anomalous jet.
This illustrates how generally CWola can be used. The only requirement is that one is able to define regions of the data depleted and enriched in signal, and that the signal and background events are statistically equal in the two regions. In terms of model independence, some degree of signal assumption is needed in order to define appropriate mixed samples.

Methods can also be used to bootstrap CWola and further improve the classification performance. In \citen{Amram:2020ykb}, a powerful and model-independent anomalous jet tagger is defined starting from the CWola hunting methodology, but defining the mixed samples for training differently. Targeting signals where both jets in the event are anomalous, the key idea is that for a resonance decaying to a pair of anomalous jets, one can use an initial self- or supervised classifier (like an autoencoder, see Section~\ref{sec:selfsupervised}) 
to tag an event as signal-like or background-like using one jet and then use that information to construct samples for training a classifier using the other jet with weak supervision. By using an autoencoder as an initial classifier, one can group events into a signal-like and background-like sample based on the anomaly score on one of the jets (assuming that if the one jet is anomalous, the other must be too). A classifier can then be trained for the other jet that has not been tagged, where the mixed samples are defined based on the anomaly score of the tagged jet.

Another weakly-supervised method for over-density detection is ANODE~\cite{nachman2020anode}. In ANODE, conditional neural density estimation is used in order to interpolate probability densities from a data sideband into the data signal region. This interpolation is used and compared to the probability density of the actual data observed in the signal region, and used to construct a likelihood ratio as in Eq.~\ref{eq:weak-supervision-likelihood-ratio}. This implies having to learn both the interpolated likelihood of the background in the signal region, as well as the likelihood for data in the signal region (see Fig.~\ref{fig:fig1}). An improvement on this method is CATHODE (Classifying Anomalies Through Outer Density Estimation)~\cite{Hallin:2021wme}. In CATHODE, rather than directly constructing the likelihood ratio, one rather samples events from the trained background estimator after it is interpolated into the signal region. This avoids having to learn the likelihood of data in the signal region. Then, a classifier is trained to discriminate data in the signal region from the data samples from the interpolated density estimator. This algorithm was first demonstrated for searches for heavy particles decaying into two jets. CATHODE proceeds by first training a conditional normalizing flow~\cite{rezende2016variational} on the dijet invariant mass sidebands and then interpolating this into the signal region; samples from this flow are used as the background model and should correctly take into account any correlations between the input features and the dijet invariant mass.

Normalizing flows are a type of generative model that learn to transform a simple probability distribution (usually a standard Gaussian distribution) into a more complex distribution that resembles the target distribution of the data. This is achieved by defining a sequence of invertible transformations that map samples from the simple distribution to samples from the target distribution. The resulting model can be used to generate new data samples, perform density estimation, and compute likelihoods. Invertibility is important, as it ensures that the transformation has a well-defined inverse, which is needed for density estimation and likelihood computation. The key challenge in designing normalizing flows is to ensure that the resulting distribution is both complex enough to capture the target distribution and easy to work with, in the sense that likelihood computation and sampling are efficient. Recent work has focused on designing more expressive and flexible transformations, such as coupling layers, which allow for the transformation to depend on only a subset of the input variables~\cite{dinh2017realnvp}. CATHODE utilizes such a normalizing flow to estimate the background density, conditioned on the dijet invariant mass. The density can then be interpolated into the dijet invariant mass signal region, while accounting for all correlations between the input features. Finally, a classifier is trained to distinguish between the artificially generated background samples from the normalizing flow (trained in data sidebands) and actual samples from the data signal region, yielding an estimate of the likelihood ratio as an anomaly metric (following the CWola paradigm).

A similar method is CURTAINS~\cite{Raine:2022hht}. This method also takes advantage of a conditioned invertible neural network to learn the distribution of background events in a sideband and then use that to transform datapoints to those of the target distribution in the signal region. CURTAINs use an optimal transport loss to train the network to minimize the distance between the model output and the target data. The goal is to approximate the optimal transport function between two points in feature space when moving along the resonant spectrum. As a result, instead of generating new samples to create a background template, CURTAINs transforms the data in the side-bands to equivalent data points with a mass in the signal region. This approach eliminates the need to match data encodings to an intermediate prior distribution, which is the case of CATHODE, as it can lead to mismodelling of underlying correlations between the observables in the data if the trained posterior is not in perfect agreement with the prior distribution. Additionally, CURTAINs can also be employed to transform side-band data into validation regions, rather than simply constructing the background template in the signal region, making the algorithm easier to validate and test. Once the CURTAINs density estimation algorithm has been trained, a similar approach as in CATHODE is taken. Specifically, the transformed data (from sideband to signal region) is assumed to represent a sample of pure background events, while the signal region data represents a mixture of signal and background. A CWola classifier is trained to discriminate between the two datasets based on this assumption. 

In Ref.~\citen{klein2022flows,curtains2}, an improvement of the CURTAINs technique is introduced, where a maximum likelihood estimation is used instead of an optimal transport loss. This improves the fidelity of the transformed data and is significantly faster and easier to train.

More recently, diffusion models~\cite{10.5555/3045118.3045358}, emerging as potent tools for high-dimensional density estimation, have been explored both for overdensity estimation~\cite{sengupta2023improving} and for outlier detection~\cite{mikuni2023highdimensional}.

There are also weakly supervised methods that take advantage of simulation in the training of density estimators. In Simulation Assisted Likelihood-free Anomaly Detection (SALAD), a reweighting function for reweighting simulation to match data in the data sidebands is trained. This (parametrized) reweighting function is then interpolated into the signal region. Finally, a classifier to discriminate between the two is trained to get the likelihood ratio. Another simulation-assisted technique is Flow-enhanced transportation for anomaly detection (FETA)~\cite{feta}, a mixture of SALAD and CURTAINS. A normalizing flow is trained in the sideband to map MC simulation to data. This learned flow is then applied to simulation in the signal region to obtain an approximation of the background.

There are caveats when deploying weakly supervised methods.
Asymptotically, a weakly supervised classifier will converge to the performance of a fully supervised one. But in practice, performance typically degrades with smaller samples sizes available for training and lower fractions of signal events in the data sample. However, one can still obtain signal versus background classifiers with reasonable performance even with signal fractions well below 1\%. Biases in the background model can also lead to degraded performance; if the classifier is able to distinguish between the background events in the two samples it will learn to encode that difference instead of learning to identify signal events.



\subsection{Outlier detection}
The above methods focus on detecting new physics as overdensities in very specific regions of the kinematic phase space; this paradigm is similar to a traditional bump hunt, often performed in HEP searches for novel particles. However, new physics signatures are equally likely to manifest themselves as unexpected events in the tail of distributions. This type of events may be identified using out-of-distribution detection algorithms. The prime example of such an algorithm is the auto-encoder\cite{lecun1987phd, ballard1987modular, hinton1993autoencoders}, which is especially popular in high-energy physics applications\cite{Radovic:2018dip, albertsson2019machine, Jawahar:2021vyu, tsan2021particle, Finke_2021, Laguarta:2023evo, Vaslin:2023lig, Anzalone:2023ugq, Bohm:2023ihd}.

\subsubsection{Self-supervised methods}
\label{sec:selfsupervised}
\begin{figure}
  \centering
  \includegraphics[width=0.99\textwidth]{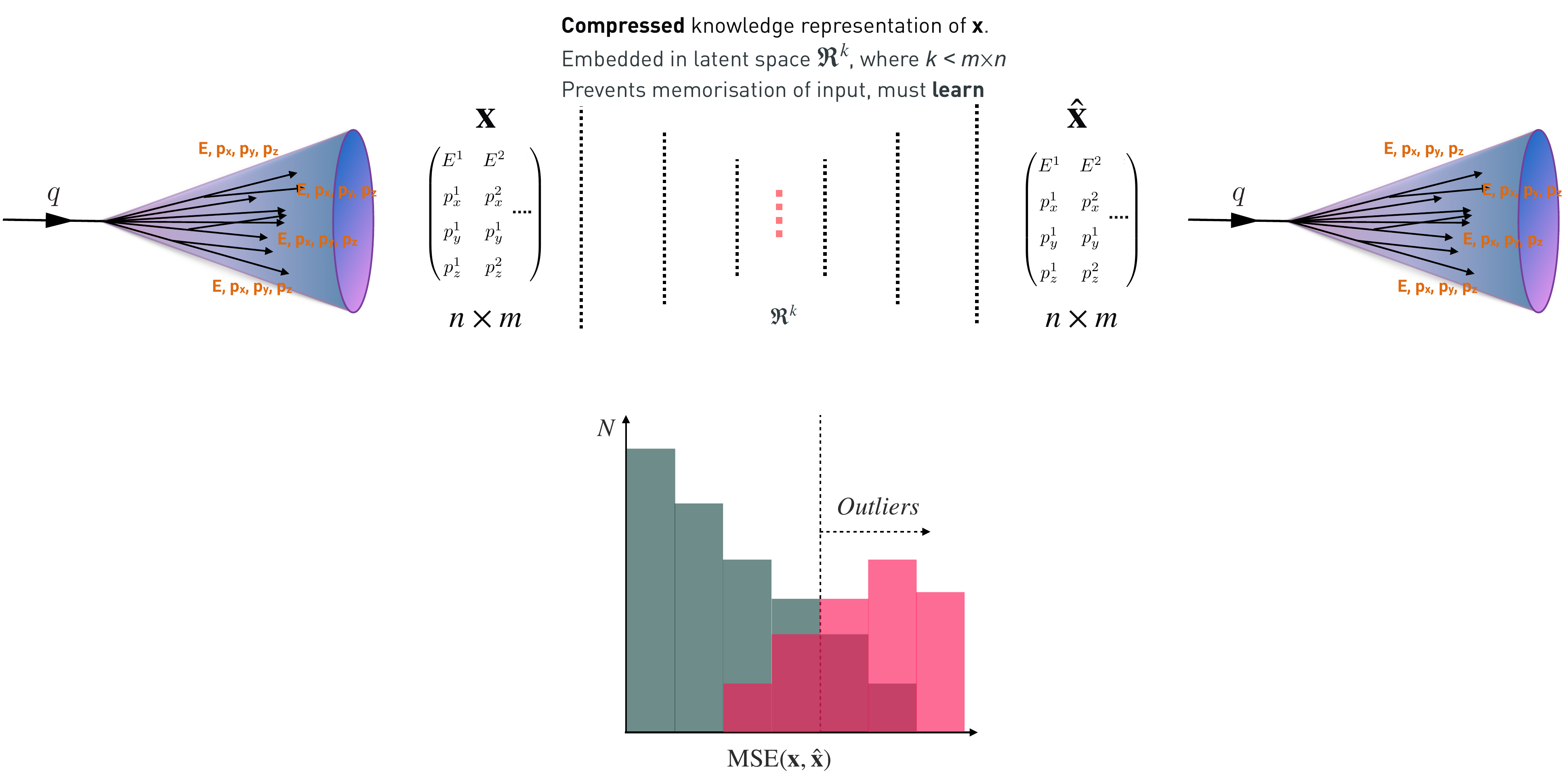}
  \caption{An auto-encoder is trained to encode the input to a lower dimensional embedded space, and then decode it again in order to reconstruct the original input (top). The difference between the input and the output can be used as an anomaly score (bottom). }
  \label{fig:ae}
\end{figure}
Self-supervised learning\cite{balestriero2023cookbook} is a form of unsupervised learning where the data provides the supervision. In general, a part of the data is initially withheld from the model and the task of the network is to reproduce this data. Consequently, the network learns a meaningful representation of the data to solve this problem. The self-supervised learning workflow usually involves two stages: first, generating a set of supervisory signals from the input data; and second, employing these signals for a supervised~task. Self-supervised learning can be seen as a hybrid approach that lies somewhere between unsupervised and supervised learning. In high-energy physics, the most used type of self-supervised model is by far the auto-encoder.

The standard Auto-Encoder (AE) model consists of two neural networks: the encoder and the decoder. The encoder maps the input data to a \textit{latent space} of a lower dimensionality. For example, a particle that is represented by 64 features (transverse momentum, azimuthal angle, etc.) is reduced to a 16 feature representation. In contrast, the objective of the decoder is to reconstruct the input features from the latent space features. The ultimate goal of the AE training is to minimise the difference between the input and reconstructed data. This difference can be quantified by employing various loss functions. The Mean Squared Error (MSE) loss function is the most basic example of quantifying the input-output discrepancy:
\begin{equation}
\label{eq:vanillaloss}
L_\mathrm{MSE} = (x - f(z,\theta))^2    
\end{equation}
where $x$ is the input data, $z$ is the latent space data, and $\theta$ are the weights of the decoder. This reconstruction loss is propagated through both the decoder and the encoder. Thus, the latent space and the reconstructed data evolve simultaneously.

The extent to which the auto-encoder latent space follows a statistical distribution is referred to as the latent space \textit{regularity}. The latent space of the standard auto-encoder does not follow any particular distribution.  The regularity of the standard AE depends on the input features, the dimension of the latent space, and the encoder architecture. Thus, the encoder will shape the latent space such that it facilitates the reconstruction task, thus minimising the MSE loss from \autoref{eq:vanillaloss}. In contrast, the Variational Auto-encoder (VAE)\cite{kingma2014autoencoding} is an extension of the conventional auto-encoder described above, which models the latent representation to approximate a given probability distribution. This is typically a Gaussian distribution, described by a mean and a variance; however, many alternatives exist\cite{joo2019dirichlet, patrini2019sinkhorn, Cerri_2019, Dillon_2021, Cheng_2023}, and the choice of latent space distribution ultimately depends on the task. 

The main idea of variational inference is to deﬁne a parametrised family of distributions and to search within it for the best approximation of the chosen prior distribution. The ``best approximation'' is defined as the element of the aforementioned family of distributions that minimises a pre-deﬁned function that measures the dissimilarity between the trial approximation and the prior. The function that is most commonly employed for this task is the Kullback-Leibler\cite{Joyce2011} (KL) divergence, defined as
\begin{equation}
    \mathrm{D_\mathrm{{KL}}}(\vec{\mu}, \vec{\sigma}) = -\frac{1}{2}\sum_i \left ( \log(\sigma_i^2) - \sigma_i^2 -\mu_i^2 +1 \right)~,
    \label{eq:D_KL}
\end{equation}
for the specific case of comparing a parametrised Gaussian distribution $\mathrm{N}(\vec{\mu},\vec{\sigma})$ with $\mathrm{N}(1, 0)$. A broader discussion on the KL divergence is found in Ref.\,\citen{paisley2012variational}. Note that the KL divergence is a somewhat unstable dissimilarity metric. Hence, more robust alternatives exist, such as the Wasserstein distance, which led to the creation of an AE architecture with the same name\cite{tolstikhin2019wasserstein}. Variations on the Wasserstein AE have also been applied in a high-energy physics context\cite{Komiske_2019, Komiske_2020}.

The VAE loss consists of two components: the reconstruction loss, conventionally the MSE, and the KL divergence term. The latter encourages the VAE to produce a latent space that follows a well-defined prior distribution, regularising the model. Thus, the VAE loss can be written schematically as
\begin{equation}
    {\cal L} = (1-\beta) \mathrm{MSE}(\mathrm{Output}, \mathrm{Input}) + \beta \mathrm{D_\mathrm{{KL}}}(\vec{\mu}, \vec{\sigma})~,
    \label{eq:vae_loss}
\end{equation}
where MSE labels the reconstruction loss, $\mathrm{D_\mathrm{{KL}}}$ is the KL regularization term, and $\beta\in[0, 1]$ is a hyperparameter that balances the effect of the two loss components.

\begin{figure}
  \centering
  \includegraphics[width=0.49\textwidth]{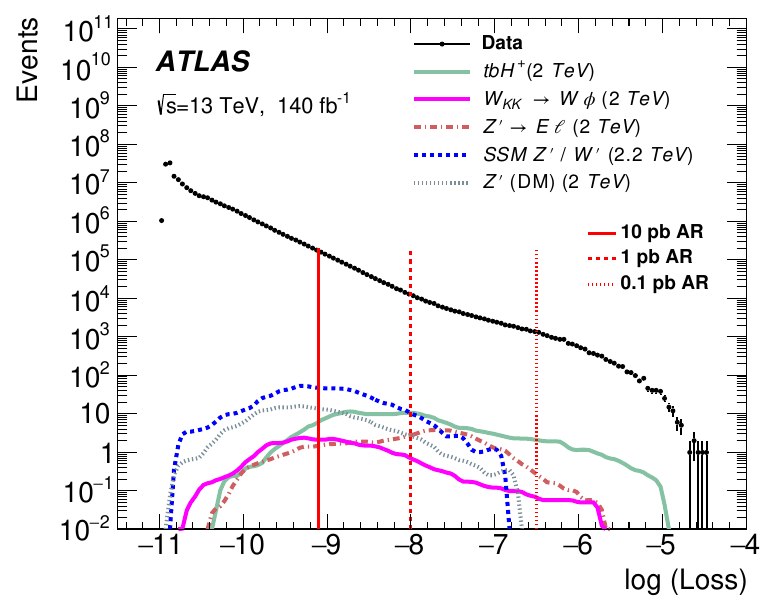}
  \caption{The output, or anomaly score, of an auto-encoder trained on data collected by the ATLAS Experiment and evaluated on data (black) and a range of benchmark BSM signals~\cite{ATLAS:2023ixc}.}
  \label{fig:atlasae}
\end{figure}

The weakly supervised methods from the previous sections aim to learn the likelihood ratio and thus can identify anomalies. In contrast, self-supervised models only learn the probability density of the background. Hence, an event may be labeled as anomalous if its probability to be associated with the learned latent distribution is very low. Additionally, the learned distribution exists in a lower dimensional embedded space. This stops the model from memorizing the input and is a form of lossy compression. Therefore, the model is generally capable of reconstructing events it is frequently exposed to during its training, but it fails at reconstructing events that are rare in the training set. The difference between the input data and its reconstructed counterpart may then be used to define an anomaly score: a high MSE is expected for anomalous data and a low MSE is expected for typical events. An illustration of this paradigm is shown in Figure~\ref{fig:ae}. There exist several studies in HEP where AEs and VAEs are used for detecting new physics as outliers in the data~\cite{Farina:2018fyg, Heimel:2018mkt,Blance:2019ibf,Hajer:2018kqm,Roy:2019jae,Cheng_2023}. For~example, this type of workflow was used to search for new physics in the two-body invariant mass spectrum of two jets or a jet and a lepton with the ATLAS Experiment in Ref.~\citen{ATLAS:2023ixc}. Therein, a selection on an auto-encoder output is used to suppress the background and define signal regions with a high signal-to-background ratio. The auto-encoder output for data and for a range of potential new physics signatures is shown in Figure~\ref{fig:atlasae}.

As mentioned in the beginning of this section, autoencoders are efficient for event-by-event outlier detection and are not expected to perform well in finding overdensities. This makes them complimentary to the weakly supervised methods. Furthermore, an additional problem that auto-encoders have is discussed in Ref.~\citen{obstructions}. In the aforementioned work it is demonstrated that the connection between large MSE and anomalies is not completely clear: for data sets with a nontrivial topology, there will always be points that wrongly are classified as anomalous. Conventionally, this can be mitigated by using VAEs and classifying anomalous events using the regularized latent space. An alternative method of circumventing this issue is based on the so called normalised AE \cite{yoon2023autoencoding}, which is located at the boundary between self-supervised and unsupervised learning. This newer type of AE architecture uses energy-based models as an alternative to the likelihood ratio or the MSE. Thus, the normalised AE avoids classifying genuinely complex albeit standard events as anomalous. For more details on this last kind of AE and its possible application to HEP, see Ref.\,\citen{dillon2023normalized}. As mentioned earlier, diffusion models are also being explored as an alternative method to perform density estimation, similar to variational autoencoders, utilizing the learned density as a permutation-invariant anomaly detection score ~\cite{mikuni2023highdimensional}.

As mentioned earlier, diffusion models are also increasingly being investigated as an alternative approach for density estimation. This method parallels the use of variational autoencoders, leveraging the learned density to create a permutation-invariant score for anomaly detection, as detailed in Ref.~\citen{mikuni2023highdimensional}.

\subsubsection{Unsupervised Methods}
\label{sec:unsupervised}

\begin{figure}
  \centering
  \includegraphics[width=0.99\textwidth]{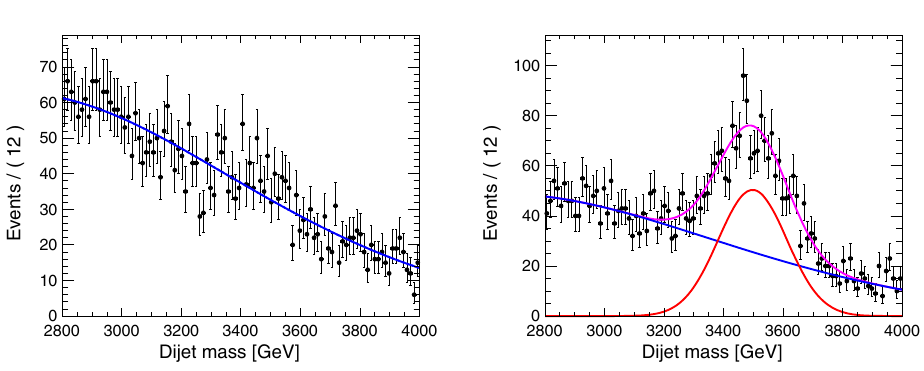}
  \caption{Dijet mass distribution for the cluster with the highest signal-over-background ratio (right) and for the cluster closest to it (left). The background component in the signal region can be modeled from the closest cluster~\cite{ucluster}.}
  \label{fig:ucluster}
\end{figure}

Unsupervised anomaly detection methods usually perform some type of data clustering. They include models such as Support Vector Machines~\cite{boser1992training}, Isolation Forests~\cite{isoforest}, and Gaussian Mixture Models\cite{vanBeekveld:2020txa, Kuusela_2012}. An application using SVMs for anomaly detection in particle physics is discussed in Section~\ref{sec:qml}. An example of unsupervised clustering for collider physics is presented in Ref.~\citen{ucluster}. Therein, the Unsupervised Clustering algorithm, or UCluster, uses an attention-based Graph Neural Network known as "ABC net"~\cite{abcnet} to create a latent space in which points sharing similar properties are placed close to each other. This is achieved by combining a clustering objective and a classification task during training. The produced embedding is shown to be capable of clustering together events that contain a new physics signal. A benefit of this method is that it naturally provides a way of performing background estimation. For each identified cluster, the nearest cluster within the embedding space can be used as a background model. The anomalous signal remains
localized in a particular cluster. Therefore, the nearest clusters are signal free, as shown in Figure~\ref{fig:ucluster}. 

A model-independent search method, based on Gaussian Mixture Models (GMMs), is introduced in Ref.~\citen{Kuusela_2012}. Within this methodology, a GMM is being used to model the background. Then, in order to avoid any dependence on a signal hypothesis, deviations from this model are identified by fitting a mixture of the background model and a number of additional Gaussians to the observed data. This allows to search for any potential deviation from the background expectation without developing a model for the signal a priori.

Finally, decision trees have also been explored in anomaly detection for searches. For example, in Ref.~\citen{roche2023nanosecond}, a tree-based autoencoder is trained through a self-supervised paradigm on background data and then evaluated on the ADC challenge data~\cite{adcchallenge}. Their unsupervised counterpart, isolation forests, have been less prominent in particle physics, but they have been applied for accelerator control~\cite{Halilovic:2665985}.

A key challenge with outlier detection methods, as discussed in Ref.~\citen{golling2023massive}, is their tendency to generate anomaly scores closely correlated to the variable of interest. This may lead to undesired sculpting effects, complicating bump-hunt like searches. To address this, strategies such as decorrelating the latent space from the variable of interest or tailoring the anomaly metric to be conditional on the jet mass \cite{Cheng_2023} should be explored. However, efforts in these areas remain limited.

\subsection{Parametrizing the alternative hypothesis}

\begin{figure}
  \centering
  \includegraphics[width=0.79\textwidth]{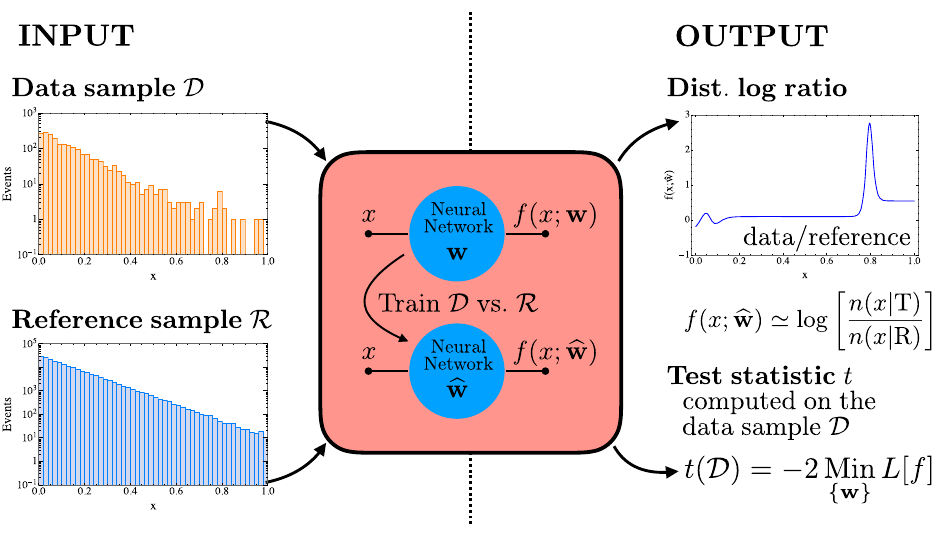}
  \caption{A schematic of the New Physics Learning Machine. Given a data sample and a reference sample, the neural network itself parametrizes the alternative hypothesis. The output of the algorithm is the ratio between the best-fit data distribution and the reference distribution, and a test statistic variable \textit{t}.~\cite{nplm}.}
  \label{fig:nplm}
\end{figure}

For true model-independent searches, no assumption should be made about the alternative model. In Ref.~\citen{Kuusela_2012}, the alternative hypothesis is parametrized as a mixture of the background model and a number of additional Gaussians. Another method is the New Physics Learning Machine (NPLM)~\cite{nplm}. Here, the alternative hypothesis is being parametrized by the network itself: given a dataset and a reference sample (like Monte Carlo simulation or data from a data sideband), a neural network is constructed such that it parametrises the alternative model as small perturbations away from the reference. When this model is trained, it learns the maximum likelihood fit to the data by construction, since its loss incorporates the log likelihood of the data. Its output is the ratio between the best-fit data distribution and the reference distribution, which is used as a test statistic to select data that displays a high level of discrepancy with the reference model. This ratio measures the disagreement between the reference model and the data and can be used for hypothesis testing. An overview of the NPLM design in shown in Figure~\ref{fig:nplm}. A drawback of this method is the difficulty of defining the reference sample $\mathcal{R}$. For example, $\mathcal{R}$ can be a taken from Monte Carlo simulation, with the caveat that this might be a less than optimal approximation to nature. Alternatively, the reference sample can be taken from a data sideband. However, in this case the difficulty is to find a region that is signal free, but still statistically identical to the data signal region. 

Integrating NPLM with techniques such as CURTAINs or CATHODE offers a potential method for creating the reference sample. This entails training a conditional density estimator with data from signal-free sidebands, enabling effective extrapolation into the signal region.
For the technique to be effective and avoid generating false positives, it's crucial that the density estimation maintains a high degree of accuracy throughout the entire spectrum of the variable of interest. One challenge arises when integrating NPLM in its full power, which is capable of identifying overdensities across multiple dimensions simultaneously, with conditional density estimation. This integration demands conditioning on multiple variables at the same time, adding a layer of difficulty to the process.

\noindent

\begin{figure}[bth]
  \centering
  \includegraphics[width=0.99\textwidth]{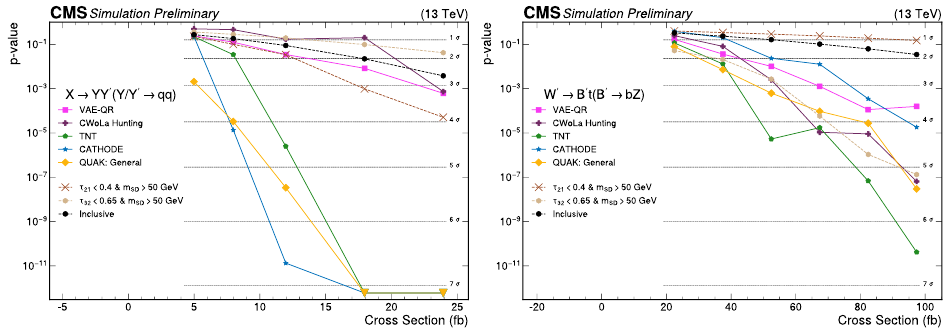}
  \caption{p-values as a function of injected signal cross sections for different anomaly detection methods (solide lines) for two different signals: X→YY’ with $M_X= 3$~TeV, $M_Y= 170$~GeV and $M_{Y’}= 170$~GeV (left), and W’ →tB’ with $M_{W’}= 3$ TeV and $M_{B’}= 400$~GeV (right). Several reference methods (dashed lines) are also added: an inclusive dijet search (black), a traditional two-prong-targeted event selection (brown), and a traditional three-prong-targeted selection (beige).~\cite{Harris:2881089}.}
  \label{fig:pvals}
\end{figure}

A challenge in utilizing anomaly detection for discovering new physics lies in the inherent difficulty of optimizing these algorithms when the nature of the signal remains unknown a priori. Moreover, the sensitivity of various anomaly detection methods can vary considerably depending on the type of signal, as demonstrated in Figure~\ref{fig:pvals} and elaborated in Ref.~\citen{Harris:2881089}. The best one can hope to do is to monitor the performance on wide variety of different potential signals 

\section{Real-time anomaly detection}
\label{sec:realtime}
The ultimate limiting factor for many searches for new physics is the event selection system of particle detectors. Tens of terabytes of data per second are produced from proton-proton collisions occurring every 25 ns, an extreme data rate that can not be read out and stored. The rate is reduced by a real-time, two-stage event filter processing system -- the {\em trigger} -- which decides whether each collision event should be kept for further analysis or be discarded. The first stage, the Level-1 trigger, is completely hardware-based running around one thousand large field programmable (FPGA) gate arrays on custom boards. The data is buffered close to the detector while the processing occurs, with a maximum latency of ${\mathcal O}(1)~\mu$s to make the trigger decision. High selection accuracy in the trigger is crucial in order to keep only the most interesting events while keeping the output bandwidth low, reducing the event rate from $40$~MHz to $100$~kHz. The data accepted by the Level-1 trigger, are read out from the detector and sent to the second software-based event filtering system, the High Level Trigger (HLT). Here the data rate is further reduced from $100$~kHz to $1$~kHz. This processing is done on commercially available CPUs and, recently, also on GPU accelerators~\citen{Bocci_2023}. The latency requirement at the HLT is ${\mathcal O}(100)$ms.

Recently, it has been proposed to look for Beyond Standard Model physics signatures in a model-agnostic way at the trigger level, both at the HLT~\citen{vaemining} and at the Level-1~\citen{ad_nmi} stage. The existing selection algorithms within the trigger system currently prioritize collisions that generate high-energy outgoing particles. However, these algorithms have reduced sensitivity to e.g. signatures involving a high multiplicity of low-momentum particles. To address this, outlier detection techniques have garnered attention for their potential to enhance acceptance rates for events that are challenging to capture using conventional algorithms. The challenge when designing such models is to adhere to the strict latency, resource and throughput constraints of the trigger.

Algorithm targeting the completely hardware-based Level-1 system are especially difficult to design. Deploying ML algorithms on FPGAs presents a significant challenge due to the specialized engineering expertise required. Unlike traditional software implementations, which can be executed on general-purpose processors, FPGAs demand a deep understanding of hardware design and optimization. The process of mapping complex mathematical operations, like those found in neural networks, onto FPGA circuits is intricate and requires careful consideration of factors such as data flow, parallelism, and memory access patterns. This becomes especially important in the Level-1 trigger, where the maximum latency per algorithm can be as low as 50 ns and only a few percent of the FPGA resources can be allocated to one specific algorithm. Recently, this process has been made easier through the introduction of software libraries that perform an automatic translation of ML models into highly parallel FPGA firmware, hls4ml~\cite{hlsfml} and Finn~\cite{finn}. These libraries are interfaced to libraries that perform \textit{quantization-aware training}, a method for reducing the numerical precision of weights and activations in a neural network during training, hence reducing their memory footprint.

Utilizing these tools, Ref.~\citen{ad_nmi} demonstrates that real-time anomaly detection using a variational auto-encoder architecture is feasible within 100 ns and using only a fraction of the FPGA resources. This is made possible through quantization-aware training, and clever architecture choices. For instance, rather than using the mean-squared error between the input and the output of the autoencoder as anomaly score, only the KL divergence term entering the VAE loss is used. The benefits of using the encoder and the KL term only is that one can avoid performing Gaussian sampling on the hardware, saving resources and latency by not having to evaluate the decoder and in addition there is no need to buffer the input data for computation of the MSE.
This demonstration of the capability to perform real-time anomaly detection on FPGAs has generated attention within the community and initiated a challenge similar to those found on Kaggle, the ADC 2021 challenge~\cite{adcchallenge}, as well as a dataset~\cite{ADC} for benchmarking such algorithms.

\begin{figure}
  \centering
  \includegraphics[width=0.99\textwidth]{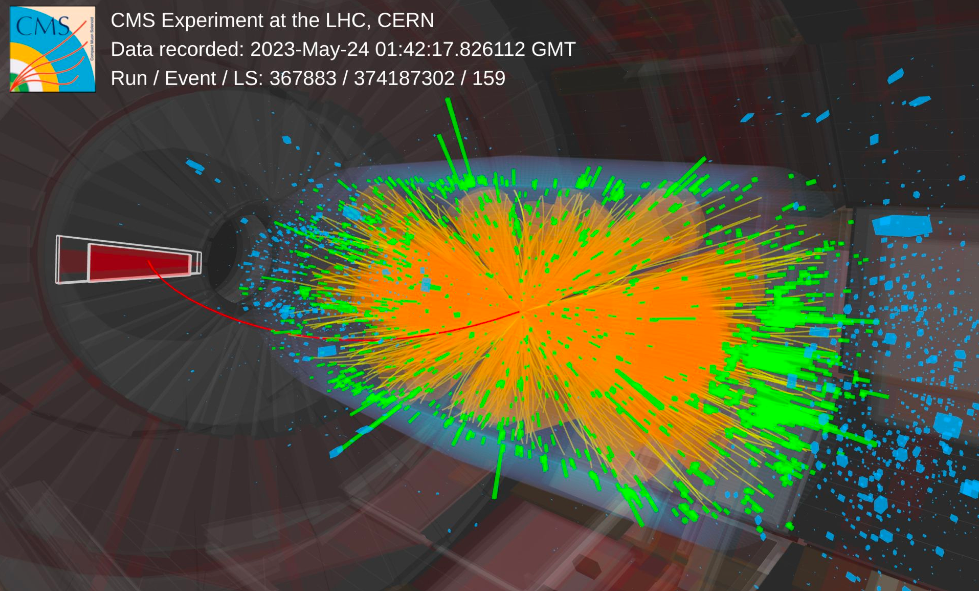}
  \caption{An event selected by an autoencoder-based anomaly detection hardware triggering algorithm in the CMS Experiment (figure from Ref.~\citen{axo}).}
  \label{fig:axo}
\end{figure}

Recently, an outlier detection algorithm similar to the one described in Ref.~\citen{ad_nmi} has been deployed into a copy of the CMS Experiment Global Trigger (GT) board~\cite{axo}. This copy of the GT system receives exactly the same input as the CMS GT, but cannot trigger a full read-out of the CMS detector, making it an excellent test-bench for algorithms targeting the main system. The anomaly detection algorithm, referred to as AXOL1TL, has been trained on unbiased data collected by the CMS experiment and shown to improve the signal efficiency for a range of different BSM signals by up to 46\%, without significantly increasing the background rate. Taking as input a subset of the available information available in the CMS Level-1 Trigger (the four-momentum of 10 jets, 4 muons, 4 electrons/photons and missing energy), and returning an output anomaly score for each event, AXOL1TL operates at an extremely low latency of 50 ns and uses less than 1\% of the FPGA resources. This is made possible through aggressive quantization and parallelization of the autoencoder.
Figure~\ref{fig:axo} shows an event display of the highest anomaly score event selected by AXOL1TL after analyzing CMS data collected in 2023. The event does not pass and other L1 trigger algorithm and is characterized by a very high number of low to medium momentum jets.

When using outlier detection as a way of discovering new physics phenomena, the detection of outliers is not sufficient. A full statistical framework for hypothesis testing is also necessary in order to claim discovery. This requires a background estimate with which the observed data in the signal region can be compared to, and this can be either based on simulated data or it can be fully data-driven. This is also true when it comes to analyzing data collected by an \textit{anomaly trigger}. From the discussion above on density estimation, the background estimate was an important part of the anomaly detection method itself, for instance in the CWoLa bumphunt the estimate was taken from the regions adjacent to the signal region and in Cathode it was sampled from the ML-generated density estimate itself. For autoencoder-based anomaly, the background estimate is not provided by the model itself. It only offers a method for achieving high signal sensitivity. In Ref.~\citen{PhysRevD.105.055006}, a statistical method for detecting non-resonant anomalies using auto-encoders is introduced. In this approach, multiple autoencoders are trained with the aim of maximizing their independence from one another. This is achieved by utilizing the distance de-correlation (DisCo) method~\cite{PhysRevLett.125.122001,PhysRevD.103.035021}, where a regularizer term based on the DisCo measure of statistical dependence is included in the training. Events are classified as anomalous if their reconstruction quality is poor across all autoencoders. Instances classified as anomalous by one autoencoder, but not by all, provide the necessary context for estimating the Standard Model background in a manner that is not model dependent. This estimation is carried out using the ABCD method. The ABCD method is commonly used for data-driven background estimation in particle physics. It consists of designing four data regions, A, B C, and D, based on orthogonal selections on two independent variables, $\tau_1$ and $\tau_2$, and then transferring the background prediction from the three signal-free regions into the signal region. In Ref.~\citen{PhysRevD.105.055006}, the two variables are the anomaly scores of the two autoencoders, which are now statistically independent due to the inclusion of the DisCo term at training time. In order to use this to design an anomaly detection method that can run in the trigger system, the authors propose to preserve all events falling within the signal-sensitive region as defined by the two autoencoders. Additionally, a random subset of events in the other three regions would be conserved for subsequent offline background estimation.

Trigger algorithms typically focus on achieving a specific signal efficiency. However, in anomaly detection, which aims to be sensitive to a broad spectrum of new signals, there isn't a definite signal efficiency to target. Instead, these algorithms are calibrated for a certain false positive rate (FPR), which is directly related to a predetermined trigger rate. This rate generally falls within the range of around $\mathcal{O}(10-100)$~Hz, necessitating an FPR near $\sim10^{-6}$. During the tuning process, the signal efficiency across various potential new physics scenarios is tracked for guidance, along with the loss on the self-supervisory label (measured as the Mean Squared Error between input and output). Yet, developing improved metrics for refining outlier detection methods, particularly those that include sensitivity to novel, unidentified signals, remains a challenging and unresolved task.

\section{Anomaly detection for detector monitoring}
\label{sec:dqm}
Anomaly detection can also be used for automatic data quality control in particle detectors.

Data quality control involves measuring physical properties of proton collision products, and attempting to detect irregular behavior within subdetectors. This can for instance be cases where a portion of the subdetector becomes unresponsive. When such anomalies occur, their impact is manifest in the properties that are measured or reconstructed from the data.

To maintain high data quality, data quality monitoring (DQM) traditionally relies on some predefined set of statistics and rules that define the expected or normal values for these statistics. An important feature of these metrics is their capacity to flag anomalies when significant deviations from the expected distributions are detected. The creation and maintenance of such statistics require deep understanding of the detector and the potential anomalies it may encounter.

As an alternative, ML based solutions have been proposed to automate this process~\cite{Borisyak_2017, Pol:2790963,pol2018detector}. In Ref.~\citen{Pol:2790963}, a thorough comparison of supervised, semi-supervised and self-supervised approaches for the detection of potential detector failure is introduced. For a classifier based algorithm, the caveat of the necessity of obtaining a large amount of training data is overcome by using a low number of training parameters in the model. To improve the anomaly detection capabilities of autoencoders used for DQM, a mix of labelled negative and unlabelled samples are used, so-called semi-supervised novelty detection.
 Autoencoder-based anomaly detection has been integrated into the CMS Experiment online DQM infrastructure~\cite{pol2018detector}. For the reasons discussed in Ref.~\citen{obstructions}, autoencoders are not necessarily the best architecture for such tasks, as the outliers one are attempting to identify might be a simpler subset of the training data and will hence not be classified as outliers. Again, here variational autoencoders and classification in the latent space can help.

\section{Quantum Anomaly Detection}
\label{sec:qml}
\subsection{Background}
In recent years, \ac{QML} has emerged as a new paradigm for data processing at the intersection of machine learning and quantum information processing. Quantum computing has the potential to address real-world challenges that are difficult or even intractable for classical computers~\cite{arute_quantum_2019, zhong_quantum_2020, madsen_quantum_2022}. 
Such problems include prime number factoring~\cite{shor97}, a problem at the basis of classical modern-day encryption, search in unstructured databases~\cite{grover1996fast}, solving systems of equations~\cite{Harrow_Hassidim_Lloyd_2009}, and simulations of quantum systems, enabling first-principle computation of chemical properties in atomic, molecular, and nuclear systems~\cite{Kandala2017, Barkoutsos2018, Kiss2022}\footnote{Please also refer to references therein, references provided are non-exhaustive.}.

Initially, applications of quantum computing in \ac{ML} focused on investigating speedups in computationally expensive subroutines of learning algorithms, such as optimization and matrix inversion~\cite{Lloyd_Mohseni_Rebentrost_2014, Wiebe_Braun_Lloyd_2012, Rebentrost_Mohseni_Lloyd_2014}. Through this scope, replacing classical subroutines with quantum algorithms provide provable speedups in terms of runtime complexity of the \ac{ML} training. Nevertheless, such proofs frequently require large and fault tolerant quantum hardware. Namely, quantum computers with error correction schemes that are able to arbitrarily suppress the inherent logical error rates~\cite{Acharya2023}. Such devices do not exist yet. Currently available quantum computers are noisy and have limited number of qubits of small decoherence times~\cite{preskill_nisq_2018}. Hence, the size of the quantum circuits and the number of operations that can be carried out at present are limited. Quantum algorithms with too many operations for the device at hand, can be rendered useless, or at least equivalent to a classical computation, by the inherent hardware noise.

Lately, studies have also investigated the potential of quantum computing to enhance fundamentally the learning model itself~\cite{Biamonte2017, Schuld2018, Schuld_2018_book, Havlicek2019}. \ac{QML} models have been shown to generalise well with few training data~\cite{caro_generalization_2022}, to provide advantages over classical algorithms for specific types of learning problems~\cite{Liu2021rigorous, Kubler2021, Huang2021, pirnay22_superpol, muser2023}, and are able to identify patterns in data that cannot be  recognised efficiently  with classical methods~\cite{huangQA2022}. Mirroring classical models for classification tasks, \ac{QML} algorithms can be coarsely grouped into two categories: \textit{kernel-based} methods and \textit{variational} learning approaches\cite{Cerezo2022}.

In the former category a main example are \ac{QSVM}, where a classical \ac{SVM}~\cite{svmVapnik} is equipped with a \textit{quantum kernel}. The values of the kernel are evaluated on a quantum computer through measurements~\cite{Schuld_2019, Havlicek2019}. During training, (Q)SVMs find the hyperplane that separates different classes of data, maximizing the margin between them. These models can create non-linear decision boundaries in the data input space when equipped with kernel functions~\cite{svmVapnik}. The kernels are constructed by feature maps that transform the data into a higher dimensional feature space, in which the classes can be more effectively separated by a linear decision boundary. In the case of a quantum kernel, the data is mapped to the Hilbert space spanned by the qubit states. The dimensionality of this space grows exponentially with the number of qubits, and hence, such models are difficult to simulate classically. After constructing the quantum kernel matrix from the measurements the loss function of the model is minimized on a classical device using quadratic programming techniques. In particle physics, SVMs can be used for supervised classification tasks~\cite{vaiciulis2003, sforza2013, sahin2016}, although their use is not as prevalent as deep learning approaches or ensemble models such as boosted decision trees. Additionally, kernel machines have been extended to an unsupervised setting~\cite{one_class_svm}, where the training data is assumed to contain mostly background events and an upper bound on the expected anomaly contamination is set using a hyperparameter.


The latter category encompasses parametrised quantum circuits, also referred to as \ac{QNN} or variational quantum circuits. These circuits are composed of gates whose parameters can be tuned iteratively to minimize a loss function using classical gradient-based learning techniques~\cite{Benedetti2019, Mitarai2018}. The output of the circuit is an expectation value of an operator, that is sampled from a quantum computer. This approach allows for the training of quantum circuits to perform specific tasks, such as classification and generative modeling. Specific architectures of QNNs have been shown to be universal function approximators~\cite{Pérez-Salinas2020}, and that they can be expressed as a Fourier series expansion~\cite{Schuld_Sweke_Meyer_2021}. Contrary to (quantum) kernel machines, the loss function landscape of QNNs is non-convex, which can lead to trainability issues similar to the ones of the vanishing gradient problem in classical neural networks~\cite{Cerezo2021, Wang2021, Holmes2022}. Nevertheless, the authors of Ref.~\citen{thanasilp2022exponential} argue that under certain conditions, kernel-based models can also manifest similar problems in training. Additionally, some works provide a unified view of \ac{QML}~\cite{Jerbi2023}, while others claim that the kernel-based learning is more natural for \ac{QML}~\cite{Schuld_qml_is_kernels_2021}.

In most current applications, one can treat the quantum computer as a specialized processing unit,  that is part of the overall computation. The hybrid algorithms discussed above aim to leverage the different strengths of classical and quantum processing units while mitigating their corresponding weaknesses.

\subsection{Applications in High Energy Physics} 
Studies have assessed the potential of quantum computing and variational algorithms for simulations of lattice field theories~\cite{atas2021, mildenberger2022probing, funcke2023review}, as an alternative to \ac{MC} techniques~\cite{Kiss_MC2022, Delgado_Hamilton_2022, Chang2021, Bravo2022}, and for parton showering simulations~\cite{nachman_qshower2021, bepari2022}. Research along these lines is motivated by the question of whether quantum algorithms can provide a natural platform for simulating fundamental physics~\cite{dimeglio2023quantum}. An additional motive is the prospect of quantum computers providing a more favorable computational complexity than currently available classical methods. Furthermore, \ac{QML} models have been developed for solving reconstruction problems in the context of collider experiments~\cite{Tuysuz_2020, Grossi2020, magano2022, Lerjarza2022, duckett2022}. 

\subsection{Supervised classification}\label{sec:quantum_supervised}
In terms of classification tasks in a model-dependent setting, quantum computing was first considered in Ref.~\citen{Mott:2017}. Therein, the training of a classifier for $H\to\gamma\gamma$ events was mapped to a quantum annealing task. 
Since then, studies have mainly focused on the design and implementation of supervised \ac{QML} algorithms based on different \ac{QSVM} and \ac{QNN} architectures that are able classify \ac{HEP} events by discriminating the signal distribution from the background distribution~\cite{terashi2021, blance_quantum_2021, qmlHiggs2021, Guan_2021, Heredge2021, Chen2020, Chen2021, wu_2021_kernel, wu_2021_qnn}. Such quantum models, are often developed and assessed via computationally expensive quantum simulations on classical processors using limited number of qubits; typically up to 20. In these simulations, the algorithms can be investigated in an ideal noiseless environment. After the architecture has been chosen and its hyperparameters have converged to values that lead to good performance on the learning task at hand, the \ac{QML} algorithms are tested by running experiments on real hardware via cloud-based platforms. 

The developed quantum models are typically benchmarked against classical models of similar complexity, that are trained on the same small data sets.  So far, the number of training data points is at the order of $10^2$ to $10^4$. \ac{HEP} datasets are frequently high dimensional, with number of features exceeding the order of presently available number qubits, posing a challenge for direct input and processing by data encoding circuits on current quantum devices. To address this challenge, a set of reduced features is typically used as an input to the \ac{QML} models. This representation of reduced dimensionality is obtained by manual selection of physical variables~\cite{terashi2021, blance_quantum_2021}, Principle Component Analysis (PCA)~\cite{wu_2021_qnn, wu_2021_kernel, Schuhmacher23, Peixoto2023}, or autoencoders~\cite{qmlHiggs2021, wozniak_belis_puljak23}. In the case of autoencoders, the compression of \ac{HEP} events can be regarded as more representative since these models can, at least approximately, retain non-linear correlations of the input features in their latent space. Such higher order correlations are removed by definition in the case of PCA, and are potentially lost in manual feature selection or feature extraction based on univariate discrimination metrics~\cite{qmlHiggs2021}.

So far, in most studies regarding supervised models, the performance of the quantum algorithms is competitive and matches the performance of their classical counterparts. Numerical evidence suggests that \ac{QML} algorithms might outperform classical models when the training datasets are small~\cite{terashi2021, wu_2021_kernel, Guan_2021, Gianelle2022}. However, such a property has not been proven in general, yet.

\begin{figure}[t]
  \centering
  \includegraphics[width=0.99\textwidth]{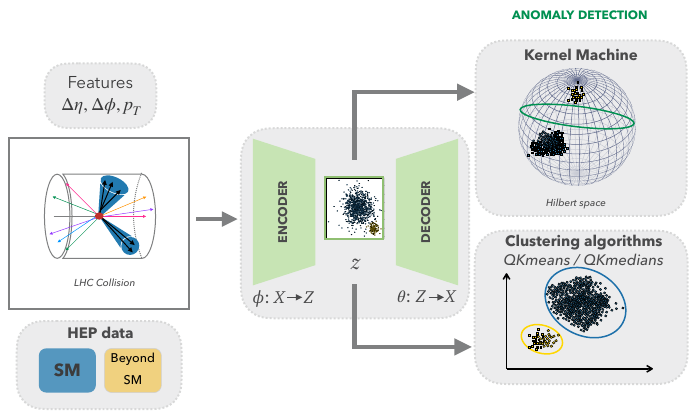}
  \caption{A typical classical-quantum pipeline. The input data is compressed using dimensionality reduction in the form of an autoencoder, and is then passed to an unsupervised quantum kernel machine and clustering algorithms~\cite{wozniak_belis_puljak23}.}
  \label{fig:qml_ad}
\end{figure}

\subsection{Unsupervised new-physics searches}\label{sec:quantum_unsupervised}
Recently, different strategies were proposed for new-physics searches at the \ac{LHC} using \ac{QML} in the context of anomaly detection~\cite{Blance_Spannowsky_2021, Ngairangbam_2022, alvi2023, wozniak_belis_puljak23, Schuhmacher23, Bermot2023, Bordoni:2023lad}. In Ref.~\citen{Blance_Spannowsky_2021}, Gaussian Boson Sampling (GBS) is used to create a lower dimensional representation of \ac{BSM} events where the Higgs boson decays into two pseudoscalar particles. GBS is classically difficult to simulate and can be implemented using continuous variable devices such as photonic quantum computers. This procedure is combined with a quantum version of the K-means clustering algorithm, Q-means, to detect anomalies. 

K-means is a method that aims to partition an unlabeled dataset into K clusters in the feature space. Each cluster center, also called a centroid, is defined as the mean of the datapoints that belong to that cluster. Each datapoint is assigned to the nearest centroid, according to some distance measure, typically the Euclidean metric, which serves us the loss function of the algorithm. The cluster assignment and the coordinates of the centroids are iteratively  updated, according to the loss minimization procedure, until the datapoints have converged to specific stable clusters. In the case of Q-means, the datapoints are embedded in the quantum Hilbert space, where the distance calculation occurs depending on the chosen quantum circuit. Additionally, depending on the design of the quantum model the minimisation of the loss can be accomplished with a quantum or classical algorithm. K-means has been applied to \ac{HEP} for jet clustering~\cite{Chekanov2006, Thaler2012, Stewart2015}, while Q-means and its variants have been applied also for unsupervised detection of new-physics events~\cite{Blance_Spannowsky_2021, wozniak_belis_puljak23}.

A quantum autoencoder (QAE) is considered in Ref.~\citen{Ngairangbam_2022} using four physical variables as an input. The authors demonstrate, both in quantum simulation and hardware, that the QAE is a promising approach for \ac{BSM} scenarios involving a resonant heavy Higgs decaying to a pair of top quarks, and a \ac{SM} Higgs decaying to two dark matter particles. The authors of Ref.~\citen{Bordoni:2023lad} employ QAEs for the detection of long-lived particles and adapt the proposed methods for execution on real quantum hardware. Additionally, different architectures of \ac{QNN} have been investigated in Ref.~\citen{alvi2023}, using low dimensional (simulated) datasets involving Higgs-like scalar particles signatures as anomalies in a semi-supervised setting. The authors do not identify any region where the tested \ac{QML} models present an advantage in performance or in terms of the needed size of the trained dataset.

Ref.~\citen{Schuhmacher23} proposes a simulation-assisted new-physics search where supervised quantum and classical \ac{SVM}s are trained on a dataset that contains \ac{SM} processes as background and an artificial set of anomalous events obtained from the \ac{SM}-distributed features as signal. Specifically, the authors generate the distributions of the signal samples by a so-called scrambling process, in which the feature distributions of the background are smeared by the normal distribution, preserving energy and momentum conservation. Furthermore, it is demonstrated that the considered models are able to generalize to real signals such as Higgs and Graviton production events.

In Ref.~\citen{Bermot2023}, a quantum Generative Adversarial Network is designed to extract an anomaly score for each data input. The authors benchmark the proposed model and verify its efficacy in data sets where they treat the Higgs boson production and Graviton production as anomalies, respectively. Additionally, generative modeling in the context of Hamiltonian learning has been investigated for semi-leptonic top and dijet event production~\cite{araz2023}. The anomaly score in Ref.~\citen{araz2023} is constructed using the different properties of the time evolution of quantum states that represent background and signal data.

A new-physics search in dijet topologies is addressed in Ref.~\citen{wozniak_belis_puljak23}, where an unsupervised quantum kernel machine and quantum clustering methods are designed to define a metric of typicality for QCD jets. The dijet events are described by 600 features --100 particle constituents per jet and three features per particle-- and the examined anomalies include two different Graviton scenarios and a \ac{BSM} scalar boson production with the final state. The authors develop a convolutional autoencoder to produce a compressed representation of the \ac{HEP} events, addressing the challenge of directly processing realistic high-dimensional data on current quantum devices. Consequently, the quantum anomaly detection algorithms use as an input the latent representation of the data that is generated by the encoder and are trained using QCD background events. For the proposed kernel-based anomaly detection model, this work identifies an advantage in performance of the quantum model over its classical counterpart.

\subsection{Discussion \& Outlook}
In \ac{HEP} applications so far, the quantum models are not designed to explicitly manifest an inductive bias towards the structure of the chosen (simulated) particle physics datasets. In the aforementioned studies, the model architectures, i.e., quantum circuits used for the implementation of \ac{QNN}s and feature maps for the kernel methods, are constructed following ansätze in the \ac{QML} literature that have desired properties such as expressiveness and hardware efficiency. 

Many \ac{QML} algorithms have been inspired by classical model architectures, such as autoencoders~\cite{Romero_2017}, convolutional neural networks~\cite{cong2019}, equivariant models~\cite{Nguyen:2022lww}, and graph neural networks~\cite{verdon2019quantum}. Despite drawing inspiration from classical models, these quantum counterparts may exhibit distinct properties and inductive biases~\cite{Kubler2021,Bowles2023}. The studies presented in Sections~\ref{sec:quantum_supervised} and~\ref{sec:quantum_unsupervised} compare the performance of their proposed models to their classical counterparts for the task at hand. However, beyond promising results in specific problems and datasets, identifying precisely in which applications QML models could provide consistent benefits such as enhancement in model performance, or computational speed-ups, still remains an open question and an active area of research. Furthermore, due to limitations in current hardware, the behavior of \ac{QML} models in the regime that is comparable to current state-of-the-art deep learning models, i.e., having millions or even billions of training samples and model parameters, is unknown.

In general, the exploration of \ac{QML} strategies for \ac{HEP} data is, at least partly, motivated by the question of whether quantum models can exploit correlations and information existing in particle physics datasets leading to advantages in performance. It is important to note, that no studies, so far, have used quantum models for supervised or unsupervised classification in real data from \ac{HEP} experiments.

The data measured by the detectors and stored for the analysis of \ac{HEP} experiments is classical. However, a quantum field theory framework is essential to predict and properly explain the outcome of such experiments. Furthermore, remnants of the initial quantum mechanical process --particle interaction-- are still present in the data. Specifically, measuring spin correlations between particles~\cite{top_correlations_CMS2019}, observing entanglement between particles produced in proton collisions~\cite{ATLAS:2023_entanglement, Cervera2017, Severi2022, Fabbrichesi2023} and violation of Bell inequalities~\cite{Fabbrichesi2021, Afik_Nova_2022, Ghosh2023} in \ac{LHC} data has been established. Measuring these first-principle quantities highlights that data from particle physics experiments cannot be described by classical local hidden-variable theories. In conclusion, the topics discussed above represent an active field of research and hold promise for classical and quantum data analysis algorithms that can enhance our ability to probe for new-physics.

\section{Conclusion and future challenges}
Anomaly detection techniques have become an integral part of modern particle physics research, and are being utilized on multiple fronts: for new physics searches, triggering and event selection, model-independent methods, detector fault and data quality monitoring, and in quantum machine learning. Automated data quality monitoring using ML-based anomaly detection methods is now being utilized in experiments, saving significant person power. This will be important as the LHC is upgraded to the HL-LHC, collecting an order of magnitude more data that will need to be validated. With the absence of any signs of new physics at particle colliders, the development of methodologies for model-independent searches and unbiased event filtering systems for the collection and subsequent analysis of particle collider data is crucial. To this end, multiple open data challenges related to anomaly detection for new physics discovery have been created~\cite{adcchallenge,lhco,darkmachines}. This has led to a substantial increase of novel methods that are being incorporated into particle physics experiments. There are still conceptual challenges related to model-independent methods, as discussed in this review article, and more research is needed to ensure these methods are sound and consistent. For example, a central problem is validating the performance of anomaly detection algorithms developed for new physics searches. The current form of validation for this type of algorithms consists of measuring the signal efficiency on a few selected benchmark simulated signal samples. This is not ideal since good performance on a simulated anomaly sample is not equivalent to good performance in a realistic setting: perhaps a true new physics sample remains elusive to an algorithm that performs extremely well on a selection of simulated new physics hypotheses. This is not the usual context for conventional anomaly detection applications where a human expert can produce an anomaly data set by explicitly labeling anomalous samples, e.g., in identifying financial fraud \cite{jiang2023weakly}. The development of a rigorous and objective metric that measures the performance of anomaly detection algorithms for new physics searches remains an open problem. Alongside this main challenge, the robustness of such algorithms against anomalies produced by detector effects and minimal correlation of the anomaly metric with a certain data feature\cite{golling2023massive} represent important issues in anomaly detection for physics that have not yet been overcome. In summary, anomaly detection in particle physics is a rapidly growing field of research, with advancements continuously being made on the algorithmic, computational, and conceptual side. Several challenges remain to be addressed, ensuring that the field will continue to grow in the coming years. 

\printbibliography

\end{document}